\documentclass[letterpaper, 10 pt, conference]{ieeeconf}  % Comment this line out if you need a4paper

\IEEEoverridecommandlockouts                              % This command is only needed if 
\usepackage{amsmath,amsfonts,amssymb, bm}
\usepackage{graphicx}
\usepackage[colorlinks=true,
citecolor=black,
linkcolor=black,
urlcolor=blue]{hyperref}
\usepackage{booktabs}
\usepackage[utf8]{inputenc}

\usepackage{algorithm}
\usepackage{algpseudocode}
\algrenewcommand\textproc{}

\graphicspath{{figures/plots/}{figures/drawings/}} 

\usepackage{subcaption}	

\usepackage[mode=text]{siunitx}	% Package for Si units
\DeclareSIUnit\noop{\relax}	% To use prefix without unit
\DeclareSIUnit{\muas}{\unit{\micro\noop}as} % Micro arcseconds
\DeclareSIUnit{\mas}{\unit{\milli\noop}as}	% Milli arcseconds
\DeclareSIUnit{\as}{as}						% Arcseconds
\DeclareSIUnit{\waves}{waves}
\renewcommand*{\(}{\left(}
\renewcommand*{\)}{\right)}
\renewcommand*{\[}{\left[}

\newcommand*{\tran}{^{\mkern-1.5mu\bm{\top}}} % Transpose sympol

\newcommand{\leg}[1]{(\textcolor{#1}{\rule[0.4ex]{3ex}{0.3ex}})} % Legend
\usepackage{arydshln}

\usepackage[nolist,nohyperlinks]{acronym}
\usepackage{etoolbox}
\makeatletter
\newif\if@in@acrolist
\AtBeginEnvironment{acronym}{\@in@acrolisttrue}
\newrobustcmd{\LU}[2]{\if@in@acrolist#1\else#2\fi}
\makeatother

\newacroindefinite{RMS}{an}{a}
\newacroindefinite{SH}{an}{a}
\newacroindefinite{AO}{an}{an}

\usepackage{tikz}
\usepackage{pgfplots}
\pgfplotsset{compat=newest}

\definecolor{SO}{RGB}		{0,   136, 255}
\definecolor{LGS}{RGB}  	{255, 103,   0}
\definecolor{NGS}{RGB}  	{0,    61, 230}
\definecolor{Mult}{RGB}		{227,  31, 255}
\definecolor{GPRoff}{RGB} 	{10,  150,  10}
\definecolor{GPRon}{RGB} 	{0,    75,   0}
\definecolor{Butter}{RGB}	{180,   0,   0}

\definecolor{Ref}{RGB}  {96,   96,  96}
\definecolor{ZusätzlPeak}{RGB}{0, 204,   0}
\definecolor{Error}{RGB}{255,   0,   0}

\definecolor{yellow}{RGB}		{255, 255,   0} % 1. yellow
\definecolor{dark yellow}{RGB}	{255, 225,   0} % 2. dark yellow
\definecolor{orange}{RGB}		{255, 191,   0} % 3. orange
\definecolor{dark orange}{RGB}	{255, 140,   0} % 4. dark orange      %% LGS Daten
\definecolor{red}{RGB}			{255,   0,   0} % 5. red
\definecolor{pink}{RGB}			{255, 105, 180} % 6. pink
\definecolor{magenta}{RGB}		{255,  30, 255} % 7. magenta          %% NGS Daten
\definecolor{purple}{RGB}		{135,   0, 135} % 8. purple
\definecolor{dark blue}{RGB}	{0,     0, 140} % 9. dark blue / navy
\definecolor{blue}{RGB}			{0,    60, 255} % 10. blue
\definecolor{light blue}{RGB}	{51,  204, 255} % 11. light blue  / sykblue
\definecolor{light green}{RGB}	{124, 252,   0} % 12. light green / lime
\definecolor{green}{RGB}		{0,   205,   0} % 13. green
\definecolor{dark green}{RGB}	{25,  100,   0} % 14. dark green

\usepackage{xspace}
\newcommand{\TT}{tip-tilt\xspace}

\title{\LARGE \bf
Modal Identification of Mirror Vibrations at the VLT using Accelerometer Data
}

\author{Pascal Jaufmann$^{1}$, Aaron Buck$^{1}$, J\"org-Uwe Pott$^{2}$ and Oliver Sawodny$^{1}$% <-this % stops a space
\thanks{$^{1}$Institute for System Dynamics, University of Stuttgart, Waldburgstr. 17/19, 70569 Stuttgart, Germany. Email: {\tt\small \{pascal.jaufmann, oliver.sawodny\}@isys.uni-stuttgart.de} }%
\thanks{$^{2}$ Max Planck Institute for Astronomy, K\"onigsstuhl 17, 69117 Heidelberg, Germany. Email: {\tt\small jpott@mpia.de} }%
}

\begin{document}
% Add Bibliography Commands for IEEEtran style
\bstctlcite{IEEEexample:BSTcontrol}

%Define acronyms
%\chapter*{\acroname}\markboth{\acroname}{}
\begin{acronym}[spaceeeeeeeeee] % set longest acronyom as option
	\acro{ESO}	{European Southern Observatory}
	\acro{ELT}	{Extremely Large Telescope}
	\acro{GTC}	{Gran Telescopio Canarias}
	\acro{VLT}	{Very Large Telescope}
%	\acro{VLTI}	{Very Large Telescope Interferometer}
	\acro{VLTI}	{\ac{VLT} Interferometer}
	\acro{UT}	{Unit Telescope}
	\acro{AT}	{Auxiliary Telescope}
	\acro{AOF}	{Adaptive Optics Facility}
	\acro{LBT}	{Large Binocular Telescope}
	
	\acro{M1}	{\LU{P}{p}rimary \LU{M}{m}irror}
	\acro{M2}	{\LU{S}{s}econdary \LU{M}{m}irror}
	\acro{M3}	{\LU{T}{t}ertiary \LU{M}{m}irror}
	\acro{M4}	{\LU{F}{f}ourth \LU{M}{m}irror}
	\acro{M5}	{\LU{f}{f}ifth \LU{M}{m}irror}
	\acro{FP} 	{\LU{F}{f}ocal \LU{P}{p}oint}
	
	\acro{MICADO}	{Multi-AO Imaging Camera for Deep Observations}
	\acro{MORFEO}	{Multi-conjugate adaptive Optics Relay For ELT Observations}
	\acro{NAOMI}	{New Adaptive Optics Module for Interferometry}
	\acro{COMPASS}  {\LU{COM}{com}puting \LU{P}{p}latform for \LU{A}{a}daptive optic\LU{S}{s} \LU{s}{s}ystem}
	\acro{KOOL}		{K\"onigstuhl Observatory Opto-mechatronics Laboratory}
	\acro{AVC}		{Adaptive Vibration Cancellation}
	\acro{ACE}		{ALPAO Core Engine}
	
	\acro{AO}	{\LU{A}{a}daptive \LU{O}{o}ptics}
	\acro{SCAO}	{\LU{S}{s}ingle-\LU{C}{c}onjugate \LU{A}{a}daptive \LU{O}{o}ptics}
	\acro{MCAO}	{\LU{M}{m}ulti-\LU{C}{c}onjugate \LU{A}{a}daptive \LU{O}{o}ptics}
	\acro{LO}	{\LU{L}{l}ow \LU{O}{o}rder}
	\acro{HO}	{\LU{H}{h}igh \LU{O}{o}rder}

	\acro{WFS}	{\LU{W}{w}ave\LU{F}{f}ront \LU{S}{s}ensor}
	\acro{WFE}	{\LU{W}{w}ave\LU{F}{f}ront \LU{E}{e}rror}
	\acro{DM}	{\LU{D}{d}eformable \LU{M}{m}irror}
	\acro{SD}	{\LU{S}{s}cience \LU{D}{d}etector}
	\acro{TTM}	{\LU{T}{t}ip-\LU{T}{t}ilt \LU{M}{m}irror}
	\acro{SH}	{Shack-Hartmann}
	\acro{SMF}	{\LU{S}{s}ingle-\LU{M}{m}ode \LU{F}{f}iber}
	\acro{SLM}	{\LU{S}{s}patial \LU{L}{l}ight \LU{M}{m}odulator}
	
	\acro{LGS}	{\LU{L}{l}aser \LU{G}{g}uide \LU{S}{s}tar}
	\acro{NGS}	{\LU{N}{n}atural \LU{G}{g}uide \LU{S}{s}tar}
	
	\acro{PSF}	{\LU{P}{p}oint \LU{S}{s}pread \LU{F}{f}unction}
	\acro{RMS}	{\LU{R}{r}oot \LU{M}{m}ean \LU{S}{s}quare}
	\acro{RMSE}	{\LU{R}{r}oot \LU{M}{m}ean \LU{S}{s}quare \LU{E}{e}rror}
	\acro{SVD}	{\LU{S}{s}ingular \LU{V}{v}alue \LU{D}{d}ecomposition}
	\acro{OPL}  {\LU{O}{o}ptical \LU{P}{p}ath \LU{L}{l}ength}
	\acro{FoV}	{\LU{F}{f}ield of \LU{V}{v}iew}
	\acro{DoF}	{\LU{D}{d}egree of \LU{F}{f}reedom}
	\acro{PtV}	{\LU{P}{p}eak-to-\LU{V}{v}alley}
	
	\acro{KF}	{Kalman \LU{F}{f}ilter}
	\acro{LQG}	{\LU{L}{l}inear \LU{Q}{q}uadratic Gaussian \LU{C}{c}ontrol}
	
	\acro{PSD}	{\LU{P}{p}ower \LU{S}{s}pectral \LU{D}{d}ensity}
	\acro{GPR}	{Gaussian \LU{P}{p}rocess \LU{R}{r}egression}
	\acro{SE}	{\LU{S}{s}quared \LU{E}{e}xponential}
	\acro{GWN} 	{Gaussian \LU{W}{w}hite \LU{N}{n}oise}
	\acro{AGWN} {\LU{A}{a}dditive Gaussian \LU{W}{w}hite \LU{N}{n}oise}
	\acro{iDFT}	{inverse \LU{D}{d}iscrete Fourier \LU{T}{t}ransformation}
	
	\acro{AR2}	{\LU{A}{a}uto-\LU{R}{r}egressive process of order two}
	
	\acro{MPIA} {Max Planck Institute for Astronomy}
	
	%\acro{EKF}{\LU{E}{e}xtended Kalman \LU{F}{f}ilter}
	%\acro{IO-linearization}{\LU{I}{i}nput-\LU{O}{o}utput \LU{L}{l}inearization}
	%\acro{KF}{Kalman Filter}
	%\acro{LS}{\LU{L}{l}east \LU{S}{s}quares}
	%\acro{LTI}{\LU{L}{l}inear \LU{T}{t}ime-\LU{I}{i}nvariant}
	%\acro{LTV}{\LU{L}{l}inear \LU{T}{t}ime-\LU{V}{v}arying}
	%\acro{MFTM}{Magic Formula \LU{T}{t}ire \LU{M}{m}odel}
	%\acro{OA}{\LU{O}{o}bstacle \LU{A}{a}voidance}
	
	%\acro{PLTM}{\LU{P}{p}iecewise \LU{L}{l}inear \LU{T}{t}ire \LU{M}{m}odel}
	%\acro{RAMPC}{\LU{R}{r}obust \LU{A}{a}daptive \LU{M}{m}odel \LU{P}{p}redictive \LU{C}{c}ontrol}
	%\acro{RLS}{\LU{R}{r}ecursive \LU{L}{l}east \LU{S}{s}quares}
	
	%\acro{SNTM}{\LU{S}{s}implified \LU{N}{n}onlinear \LU{T}{t}ire \LU{M}{m}odel}
\end{acronym}

\maketitle
\thispagestyle{empty}
\pagestyle{empty}

\begin{abstract}
Recent advances in ground-based astronomy have made it possible to create optical telescopes with primary mirrors up to \qty{40}{\m} in size.
With growing mirror diameter, the suppression of non-atmospheric disturbances becomes increasingly important.
Precise knowledge of the movement of telescope mirrors is essential for understanding and compensating for vibration-based perturbations.
A model from \acs{VLT} accelerometer data for each individual mirror is developed, while the influence of wind buffeting is accounted for by a von Karman wind model.
To describe the relevant rigid body motion, we consider the piston, tip and tilt modes of the mirrors.
The identification is validated by comparing the \acl{PSD} of the measured and identified modes.
Additionally, we assess the robustness of the approach by calculating the identification error over different sections of the data.
The study indicates that the employed methods are adequate for the identification of modal telescope vibrations.
%
%The results are an important step towards advanced model-based \ac{AO} controllers for large telescopes.
It is anticipated that said findings will serve as a significant foundation for the development of advanced model-based AO controllers for large telescopes, such as \acl{LQG}.
\end{abstract}

% Recent advances in ground-based astronomy have made it possible to create optical telescopes with primary mirrors up to 40 m in size. With growing mirror diameter, the suppression of non-atmospheric disturbances becomes increasingly important. Precise knowledge of the movement of telescope mirrors is essential for understanding and compensating for vibration-based perturbations. A model from VLT accelerometer data for each individual mirror is developed, while the influence of wind buffeting is accounted for by a von Karman wind model. To describe the relevant rigid body motion, we consider the piston, tip and tilt modes of the mirrors. The identification is validated by comparing the power spectral density of the measured and identified modes. Additionally, we assess the robustness of the approach by calculating the identification error over different sections of the data. The study indicates that the employed methods are adequate for the identification of modal telescope vibrations. It is anticipated that said findings will serve as a significant foundation for the development of advanced model-based AO controllers for large telescopes, such as linear quadratic Gaussian control.

%\begin{IEEEkeywords}
%	adaptive optics, telescope vibrations, modal identification, VLT, ELT
%\end{IEEEkeywords}

% ------------------------------------------------------------------------------------------------
\section{Introduction}\label{sec:introduction}
% ---------- Introduction and Motivaton ----------
Recent trends within the field of optical observatories include the deployment of new space telescopes and the size increase of primary mirrors of ground-based telescopes to capture more light.
Today's largest telescopes such as the \ac{VLT} have primary mirrors with diameters up to 8 meters.
The effective size of these telescopes can be increased by employing interferometers like the \ac{VLTI}.
%
%This study deals with data from the \ac{VLT} with a diameter of \qty{8.2}{m}.
%
The \ac{ESO} is currently constructing the \ac{ELT} featuring a primary mirror with a diameter of \qty{40}{m}, which upon completion will be the largest of its kind in the world.
%Currently \ac{ESO} is building the \ac{ELT} with a primary mirror of 40 meters, which will be the largest telescope in the world once finished.
%
%The increase of size comes with significant challenges.
%
However, the larger the telescope, the worse the image degradation due to the effects of disturbances~\cite{Kulcsar2012b}.
%the effects of disturbances cause blurring of the images, and the larger the telescope, the worse the image quality~\cite{Kulcsar2012b}.
%
The first major perturbation arises from the light path through the Earth's atmosphere.
To sense and correct such disturbances, the technology of \ac{AO} has been developed~\cite{Kulcsar2012a}.
It uses stars, called \ac{NGS}, in the isoplanatic region around the observed science object whose light is then detected by \acp{WFS}.
In addition, more and more artificial stars (\acp{LGS}) are deployed.
The sensed aberrations are then fed back into a control loop, which uses one (or more) \aclp{DM} that can be deformed by a number of actuators beneath its surface to mimic and mitigate the distortion of the incoming wavefront~\cite{Marquis2024}.
Another challenge is posed by the vibration of the telescope mirrors themselves.
The oscillations are largely caused by wind buffeting and mechanical machines of the telescope, such as pumps or ventilation~\cite{Kulcsar2012b}.

% ---------- State of the art ----------
Two approaches to detect these vibrations are currently being developed.
The first is to use data from the \ac{WFS}~\cite{Jaufmann2024}.
Since \acp{WFS} are installed in the light path after the mirrors, the data also provides information about the movement of the mirrors.
%
%We can distinguish these vibrations from the atmospheric influences as the turbulence is inherently low-frequency and mirror vibrations are more high-frequency.
The distinction between atmospheric influences and the mirror vibrations is possible due to the inherent low-frequency character of turbulence and the higher frequency of structural oscillations~\cite{Kulcsar2012b}.
The second method involves sensing the vibrations with acceleration sensors installed directly on the telescope mirrors~\cite{Glueck2017}. 
Their data allows for the reconstruction of mirror movements via double-integration and kinematic relationships~\cite{Gilbert2010}.

% ---------- Contribution ----------
%For this purpose this study deals with analyzing the accelerometer data of accelerometers from the \ac{VLT} for low-order perturbations of the mirrors, specifically in piston, tip and tilt.
The contribution of this study is the identification and validation of a dynamical model for low-order structural disturbances at the \ac{VLT}. Therefore, accelerometer data from three mirrors is used to obtain piston, tip and tilt information.
%
% ---------- Outline ----------
The remainder of this paper is structured as follows.
First, Section~\ref{sec:setup} provides an overview of the simulation pipeline including the \ac{VLT} optics, sensors as well as a brief description of the accelerometer data.
Thereafter, in Section~\ref{sec:model}, a model for the telescope mirror movements is derived.
In Section~\ref{sec:results}, the identification results are presented, followed by a conclusion of the study in Section~\ref{sec:conclusion}.

%------------------------------------------------------------------------------------------------
\section{Simulation setup}\label{sec:setup}
To adequately gather information about the exact mirror motion, several steps are needed.
First we discuss the structure of the \ac{VLT} with its mirrors along the optical path as well as the location of the accelerometers.
Following that, the sensor data is processed.
Lastly, the mirror motion is calculated from the sensor kinematics.

\subsection{\ac{VLT}}\label{ssec:VLT_setup}
\begin{figure}
	\centering
	\vspace{1.5mm}
	\input{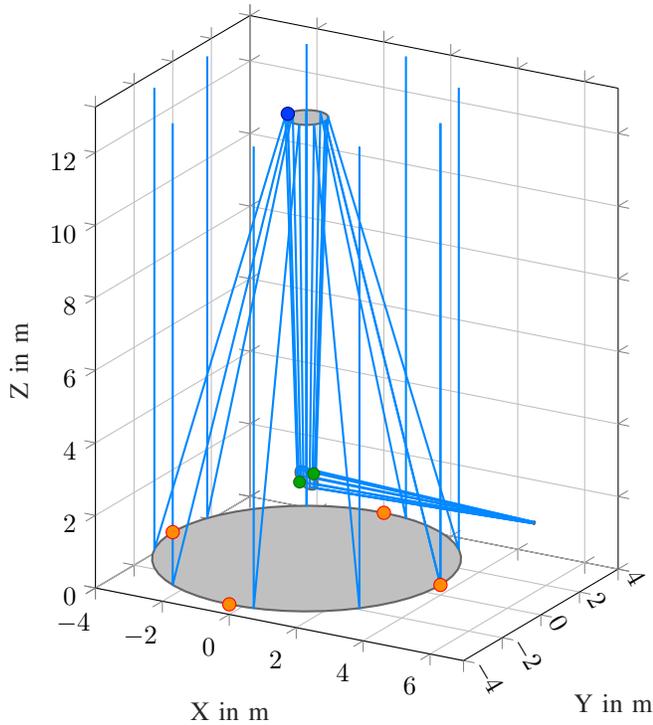}
	\caption{Optical configuration of the \ac{VLT} (Nasmyth) with accelerometer positions. Four sensors on \acs{M1} (orange dots), one sensor on \acs{M2} (blue dot), two sensors on \acs{M3} (green dots).}
	\label{fig:VLT}
\end{figure}

% --------- VLT Mirror and sensor setup ---------
The primary optical path of the \ac{VLT} consists of an arrangement of four mirrors until the first focal plane~\cite{Enard1989}.
Incoming light is first captured by the \ac{M1}, deflecting it towards the \ac{M2} at the top.
The light proceeds to the \ac{M3}, where it is redirected towards the Nasmyth \ac{FP}, or enabled to pass to the Cassegrain focus below \ac{M1}.
Depending on the science case, the light is directed to various instruments or fed via post-focal mirrors (Coud\'{e} train) to the \ac{VLTI}~\cite{Kulcsar2012b}.
In addition to wavefront sensing technology, a series of accelerometers, designed to monitor the \ac{OPL} of the four \aclp{UT}, are employed.
This functionality is crucial for ensuring that the light from the \ac{VLTI} telescopes does not suffer from destructive interference, which leads to degraded observation performance.
The specific locations of the acceleration senors are illustrated in Figure~\ref{fig:VLT}.
\ac{M1} features four accelerometers placed at equidistant points along its frame, which is sufficient for piston, tip and tilt reconstruction.
\ac{M2} has only one accelerometer mounted, while \ac{M3} features two accelerometers positioned opposite each other on the tilt axis.
The sparse distribution of accelerometers on \ac{M2} and \ac{M3} does not allow reconstruction of the piston, tip and tilt movements, required to assess the full mirror vibrations.

In this paper, we not only use the accelerometer data to determine the \ac{OPL} changes (piston) but also to capture the out-of-plane rigid body motion of the mirrors (tip and tilt).
This approach enables us to better understand and subsequently mitigate the vibrations that affect the mirrors, improving the telescope's overall stability and optical performance.

% --------- Explanation of the data we work with ---------
\subsection{Accelerometer data processing}\label{ssec:data}
\begin{figure}
	\centering
	\vspace{1.5mm}
	\input{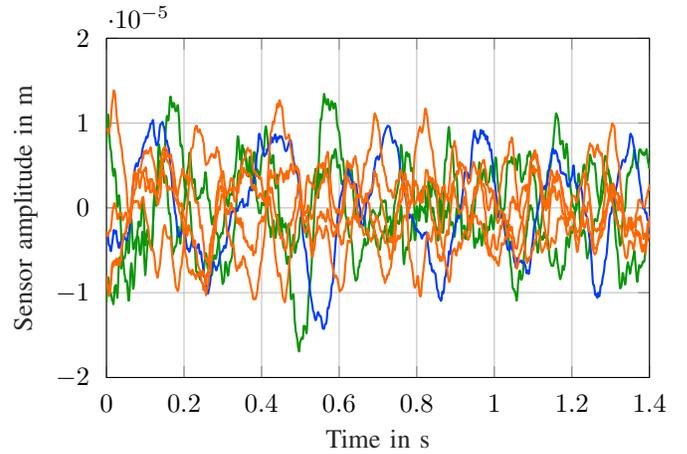}
	%	\includegraphics[width=\columnwidth]{screenshots/acc_pos1}
	%  (\tikz[baseline=-0.5ex]\draw[red, thick] (0,0) -- (0.5,0);)
	\caption{\ac{VLT} accelerometer position data. \ac{M1} sensors \leg{LGS}, \ac{M2} sensor \leg{blue}, \ac{M3} sensors \leg{GPRoff}.}
	\label{fig:acc_pos1}
\end{figure}

The uni-directional acceleration sensors are read out at a frequency of \qty{4}{\kHz}.
However, data is only logged for \qty{2.5}{\s} at the start of every minute.
This results in \num{1440} recordings a day, each \num{10.000} data points long.
The data from October 1, 2023 was kindly provided by \ac{ESO}.
To analyze only the parts of the day when the telescope is in observation mode, we extract a window of about \qty{8}{\hour} from the data.
%
%Moreover, the recordings were cut out when the telescope changed its celestial target and was therefore in rapid motion.
We cut out recordings in which the telescope changed its celestial target and was therefore in rapid motion.
Moreover, additional sections of the dataset were iteratively detected as anomalous using the identification results (c.f. Figure~\ref{fig:RMS_comp}).
The raw accelerometer signals of all 7 sensors are now integrated to obtain position information.
Forward and backward high pass filtering of the time signals is used in combination with the offline double integration.
The employed second order Butterworth filter is set to a cut-off frequency of \qty{3}{\Hz} to reduce unwanted low-frequency drifts in the data.
A small section of the retrieved sensor position signals are detailed in Figure~\ref{fig:acc_pos1}.
It is visible, that the single sensor vibrations do not exceed \qty{20}{\um}.
It seems, that two of the four accelerometers on \ac{M1} follow a similar course, whereas the other sensors deliver different signals.
The accelerometers on \ac{M2} shows a oscillation of about \qty{4.5}{\Hz} with higher frequencies embedded.

\subsection{Calculation of mirror modes}\label{ssec:mirror_modes}
\begin{figure}%
	\centering
	\vspace{1.5mm}
	%	\def\svgwidth{0.8\columnwidth}
	%% Creator: Inkscape 1.2 (dc2aedaf03, 2022-05-15), www.inkscape.org
%% Accompanies image file 'um2deg.pdf' (pdf, eps, ps)
\begingroup%
  \makeatletter%
  \providecommand\color[2][]{%
    \errmessage{(Inkscape) Color is used for the text in Inkscape, but the package 'color.sty' is not loaded}%
    \renewcommand\color[2][]{}%
  }%
  \providecommand\transparent[1]{%
    \errmessage{(Inkscape) Transparency is used (non-zero) for the text in Inkscape, but the package 'transparent.sty' is not loaded}%
    \renewcommand\transparent[1]{}%
  }%
  \providecommand\rotatebox[2]{#2}%
  \newcommand*\fsize{\dimexpr\f@size pt\relax}%
  \newcommand*\lineheight[1]{\fontsize{\fsize}{#1\fsize}\selectfont}%
  \ifx\svgwidth\undefined%
    \setlength{\unitlength}{167.43401955bp}%
    \ifx\svgscale\undefined%
      \relax%
    \else%
      \setlength{\unitlength}{\unitlength * \real{\svgscale}}%
    \fi%
  \else%
    \setlength{\unitlength}{\svgwidth}%
  \fi%
  \global\let\svgwidth\undefined%
  \global\let\svgscale\undefined%
  \makeatother%
  \begin{picture}(1,0.31357541)%
    \lineheight{1}%
    \setlength\tabcolsep{0pt}%
    \put(1.03921535,0.21502909){\color[rgb]{0,0,0}\makebox(0,0)[t]{\lineheight{1.25}\smash{\begin{tabular}[t]{c}$s$\end{tabular}}}}%
    \put(0.78389084,0.1881528){\color[rgb]{0,0,0}\makebox(0,0)[t]{\lineheight{1.25}\smash{\begin{tabular}[t]{c}$\alpha$\end{tabular}}}}%
    \put(0.0671907,0.08960655){\color[rgb]{0.14509804,0.03921569,0.02745098}\makebox(0,0)[t]{\lineheight{1.25}\smash{\begin{tabular}[t]{c}$D$\end{tabular}}}}%
    \put(0.31355633,0.22398785){\color[rgb]{0,0.03921569,0.03137255}\makebox(0,0)[t]{\lineheight{1.25}\smash{\begin{tabular}[t]{c}$\tan(\alpha) = \frac{2s}{D}$\end{tabular}}}}%
    \put(0,0){\includegraphics[width=\unitlength,page=1]{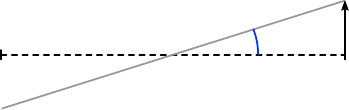}}%
  \end{picture}%
\endgroup%

	\caption{Calculation of \TT from sensor signals.}%
	\label{fig:um2deg}%
\end{figure}
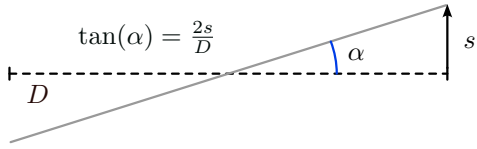

\begin{figure}
	\centering
	\input{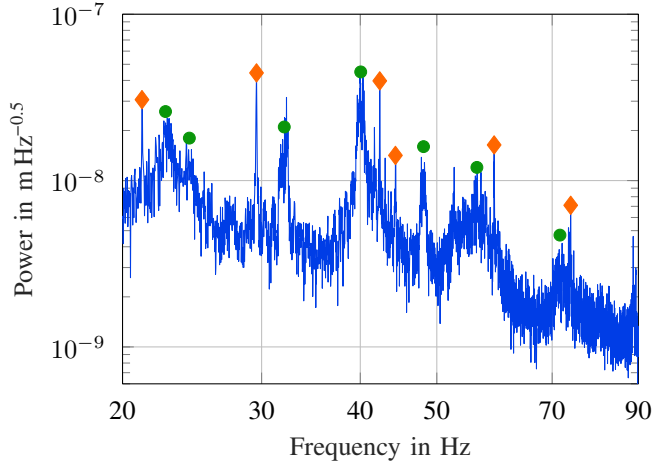}
	% Einfuegen nach xtick:
	% xticklabels={20, 30, 40, 50, 70, 90},
	\caption{\acs{PSD} of \ac{M1} piston. Eigenfrequencies of the mirror are indicated by green dots	(\textcolor{GPRoff}{\scalebox{1.2}{$\bullet$}}), other disturbances by orange diamonds	(\textcolor{LGS}{$\blacklozenge$}).}
	\label{fig:M1_piston_PSD_ex}
\end{figure}

Due to the limited distribution of accelerometers at the \ac{VLT}, we focus solely on the identification of piston, tip and tilt modes, ignoring higher-order aberrations.
In \ac{M1} and \ac{M3}, where we have more than one accelerometer mounted we calculate piston as the average of all the sensor motions.
For \ac{M2} we only have one generic mode which represents the movement of the lone accelerometer.
In this study tip represents the rotation around the x-axis and tilt the rotation around the y-axis.
Utilizing the four accelerometers mounted on \ac{M1}, we can calculate piston, tip and tilt.
For \ac{M3} we only have two sensors and thus have to resort to calculating only one mode.
As seen in Figure~\ref{fig:VLT} in the case of our installation this mode corresponds to tilt.
Tip and tilt is typically provided in \unit{arcseconds} (\unit{\arcsecond}).
For this purpose we calculate the tip or tilt in micrometers as the difference between two opposing sensors and then convert it to \unit{arcseconds} via
\begin{equation}
	\text{tip}_\text{as} = 3600\,\frac{180}{\pi}\,\alpha = \frac{648000}{\pi}\,\arctan\( \frac{2\,s}{D} \)
	\label{eq:um2as}
\end{equation}
The derivation of this conversion is visualized in Figure~\ref{fig:um2deg}.
The parameter $s$ represents the tip in \unit{\um} and $D$ the diameter of the mirror.
The angle $\alpha$ is the tip in degrees (\unit{\degree}).
For the conversion from \unit{\degree} to \unit{\arcsecond}, we need the additional factor $3600\,\frac{180}{\pi}$.
Figure~\ref{fig:M1_piston_PSD_ex} gives an exemplary temporal spectrum of the piston mode of \ac{M1} for frequencies between 20 and \qty{80}{\Hz}.
There are two distinguishable peak types seen in the spectrum that stem from two main categories of disturbances sources.
The larger and wider spots are typically eigenfrequencies from mirror support structures and the mirror itself (green dots in Fig.~\ref{fig:M1_piston_PSD_ex}).
Due to the open telescope dome, the lightweight and flexible mirror mounts are exposed to wind buffeting, exciting various frequencies in the spectrum.
The other undamped disturbance sources yield isolated peaks that have no effect on neighboring frequencies (orange diamonds in Fig.~\ref{fig:M1_piston_PSD_ex}).
These thin peaks are driven by single- or multi-frequency electrical machines such as ventilation, pumps or adaptive mirrors.
Additionally, the narrow peaks may be caused by the electrical noise of the acceleration sensors or artefacts of the position integration.
In the subsequent identification process, we focus only on the broader peaks corresponding to the mirror eigenfrequencies, as these account for most of the energy induced into the optical system.

% ---- Spektogram
% ---- Explanation of the lines in the spectogram (electric effects?)
% ---- Explanation/ Analysis of PSD (Peaks (what are actual natural frequencies, what are other effects we neglect (sharp peaks)), -40dB gradient, ...)

% ------------------------------------------------------------------------------------------------

% ------------------------------------------------------------------------------------------------
\section{Modeling}\label{sec:model}
\begin{figure}%
	\centering
	\vspace{1.5mm}
	%	\def\svgwidth{\columnwidth}
	%% Creator: Inkscape 1.2 (dc2aedaf03, 2022-05-15), www.inkscape.org
%% Accompanies image file 'spectrums.pdf' (pdf, eps, ps)
\begingroup%
  \makeatletter%
  \providecommand\color[2][]{%
    \errmessage{(Inkscape) Color is used for the text in Inkscape, but the package 'color.sty' is not loaded}%
    \renewcommand\color[2][]{}%
  }%
  \providecommand\transparent[1]{%
    \errmessage{(Inkscape) Transparency is used (non-zero) for the text in Inkscape, but the package 'transparent.sty' is not loaded}%
    \renewcommand\transparent[1]{}%
  }%
  \providecommand\rotatebox[2]{#2}%
  \newcommand*\fsize{\dimexpr\f@size pt\relax}%
  \newcommand*\lineheight[1]{\fontsize{\fsize}{#1\fsize}\selectfont}%
  \ifx\svgwidth\undefined%
    \setlength{\unitlength}{193.94640825bp}%
    \ifx\svgscale\undefined%
      \relax%
    \else%
      \setlength{\unitlength}{\unitlength * \real{\svgscale}}%
    \fi%
  \else%
    \setlength{\unitlength}{\svgwidth}%
  \fi%
  \global\let\svgwidth\undefined%
  \global\let\svgscale\undefined%
  \makeatother%
  \begin{picture}(1,0.70045325)%
    \lineheight{1}%
    \setlength\tabcolsep{0pt}%
    \put(0,0){\includegraphics[width=\unitlength,page=1]{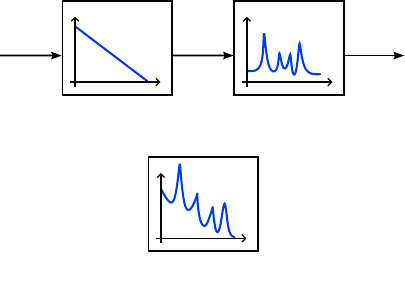}}%
    \put(0.29002867,0.41209933){\color[rgb]{0,0,0}\makebox(0,0)[t]{\lineheight{1.25}\smash{\begin{tabular}[t]{c}Wind\end{tabular}}}}%
    \put(0.7154041,0.41209933){\color[rgb]{0,0,0}\makebox(0,0)[t]{\lineheight{1.25}\smash{\begin{tabular}[t]{c}Mirrors\end{tabular}}}}%
    \put(0.50271633,0.02539464){\color[rgb]{0,0,0}\makebox(0,0)[t]{\lineheight{1.25}\smash{\begin{tabular}[t]{c}Accelerometer Data\end{tabular}}}}%
    \put(0.06573997,0.58224948){\color[rgb]{0,0,0}\makebox(0,0)[t]{\lineheight{1.25}\smash{\begin{tabular}[t]{c}GWN\end{tabular}}}}%
    \put(0.92035743,0.58224948){\color[rgb]{0,0,0}\makebox(0,0)[t]{\lineheight{1.25}\smash{\begin{tabular}[t]{c}$x$\end{tabular}}}}%
    \put(0.49884933,0.58224948){\color[rgb]{0,0,0}\makebox(0,0)[t]{\lineheight{1.25}\smash{\begin{tabular}[t]{c}$u$\end{tabular}}}}%
    \put(0,0){\includegraphics[width=\unitlength,page=2]{spectrums.pdf}}%
  \end{picture}%
\endgroup%

	\caption{The spectra of the identified wind and mirror dynamical models replicate the accelerometer data.}%
	\label{fig:spectrums}%
\end{figure}

The spectrum of the integrated measurement data exhibits a consistent negative slope, interspersed with several distinct peaks.
These peaks correspond to the various mirror eigenfrequencies, whereas the downward trend is attributable to the wind spectrum, as depicted in Figure~\ref{fig:spectrums}.
For the modeling, we assume that both pipelines are exited by \ac{GWN}.
The accelerometer data directly provides a temporal spectrum of the mirror motion~$x$ incorporating also the wind dynamics.
In our dynamical model, we first characterize the wind influence yielding the input~$u$ to the mirror transfer functions.
Afterwards, the mirror motion is isolated by extracting the relevant model parameters for the vibration-induced spectral components.
The identified mirror model is convoluted with the wind spectrum to enable a comparison to the measured accelerometer data.
To facilitate this analysis, the subsequent section details the derivation of a wind spectrum model followed by the formulation of a model representing the mirror vibrations.
Each model aims to encapsulate the respective spectral influences on the overall measurement data.

\subsection{Wind model}\label{ssec:wind_model}

To model the impact of wind forces on the \ac{VLT} mirror mounts, we utilized a von Karman wind spectrum, tailored to the Paranal Observatory conditions.
The wind force spectrum is given by
\begin{equation}\label{eq:wind_spec}
	F(f) = \frac{L_0}{V}\,\rho\,v_{\text{w,eff}}^2 \,\frac{\sqrt{\pi}}{3 \, \Gamma(2/3) \Gamma(5/6)} \frac{1}{\( 1 + \( \frac{f L_0}{V} \)^2 \)^{7/6}},
\end{equation}
where $\rho$ is the air density, $L_0$ is the outer scale of turbulence, $V$ is the mean wind speed, $f$ is the frequency, and $\Gamma$ represents the Gamma function.
The effective wind speed $v_{\text{w,eff}}$ is calculated as
\begin{equation}
	v_{\text{w,eff}} = \sqrt{\(2 V v_\text{w,rms}\)^2 + 2 \,v_\text{w,rms}^4},
\end{equation}
where the \ac{RMS} of the dynamic wind speed component is $v_{\text{w,rms}} = V I$, with $I$ being the turbulence intensity depending on the dome opening.
We follow the approach outlined in \cite{MacMynowski2010} and \cite{Bely2003}.
Later, the identified mirror model is validated by multiplying its spectrum with the wind spectrum \eqref{eq:wind_spec} and comparing the slope to the original accelerometer data.
%Once the mirror model was identified, it was validated by multiplying the identified model with the wind spectrum and comparing the resulting output to the original accelerometer data, which included the wind spectrum.

\subsection{Dynamic mirror model}\label{ssec:mirror_model}

There are different approaches to modeling a telescope and its mirrors.
A black box approach is to model the entire telescope with complex methods such as finite elements.
This and similar approaches require considerable computational effort and experimental analysis.
Additionally, models resulting from such black box methods are of very high order, making them difficult to handle and control properly~\cite{Boehm2012}.
Another method is to model the mirrors individually.
This is possible if we assume that the telescope as a whole is very stiff compared to the individual mirrors and their mounts and that there is no significant coupling between the mirrors.
These assumptions are reasonable given the huge mass of the telescope structure in relation to it's mirrors.
If we additionally neglect lateral mirror movements and solely focus on displacements orthogonal to the mirror plane, the mirrors can be represented by a mass-spring-damper system~\cite{Boehm2013}.

This study adopts a gray box modeling approach, which integrates physical modeling with data-driven parameter identification.
In particular, we employ a modal dynamical system where the respective modal parameters are estimated based on measurement data in the frequency domain.
The motions of the three relevant mirrors (\ac{M1}, \ac{M2} and \ac{M3}) consist of a number of independent oscillations (called modes), each of which is modeled as a mass-spring-damper system.
Moreover, the mirror eigenfrequencies are assumed to lie within \qtyrange{20}{90}{\Hz}.
The overall motion of the telescope mirrors is then determined by the sum of the respective mirror modes.
Specifically, the total piston, tip, and tilt of each mirror is calculated by summing the individual piston, tip, and tilt contributions from each mode.
A full model of the telescope mirror motions can thus only be achieved if we have a sufficient number of accelerometers mounted on each mirror, which is not the case for the \ac{VLT}.
While the four sensors on \ac{M1} allow for a complete calculation of piston, tip and tilt, \ac{M2} and \ac{M3} only deliver one and two modes respectively (c.f. Figure~\ref{fig:acc_pos1} and Section~\ref{ssec:mirror_modes}).
%
%Given the limitations of the available data, the most feasible approach is to approximate the overall motion of the system.
%%
%An overall mirror model is required to correct aberrations caused by mirror vibrations.
%%
%Since only \ac{M2} is deformable, compensation can only be applied at this mirror, necessitating a global model that captures the combined piston, tip, and tilt motions.
%%
%As a result, our focus is restricted to identifying the individual mirror modes.
%
The differential equation for a mirror mode is
\begin{equation}\label{eq:SMD DGL}
	\ddot{x}+2d\omega\,\dot{x}+\omega^2\,x=b\,u.
\end{equation}
The parameters $\omega$, $b$, and $d$ are identified from the \ac{PSD} of the accelerometer position data.
The frequency~$\omega$ corresponds to the location of the peak in the data, $b$ to its prominence, and $d$ to its damping.
The damping factors of the \ac{VLT} mirrors are chosen to be small due to the lightweight mirror support structures, engineered to minimize gravitational deformations and thermal expansion~\cite{Stanghellini1996}.
The \ac{M2} mounting structure at the exposed top of the telescope has been designed to avoid obstruction of incoming light and exhibits a particularly low damping.
For \ac{M1} and \ac{M3}, the damping factors are slightly higher.
This reflects the need for greater structural stability while maintaining dynamic precision~\cite{Hovsepian1998}.
%
%The damping is also chosen proportionally to the width of the corresponding peak.
%
The overall motion of the $j$-th mirror mode at the $m$-th mirror is given by the sum of all the $n_{j,m}$ individual motions.
This results in the equations
\begin{equation}\label{eq:extended SMD DGL}
	\ddot{x}_{i,j,m}+2d_{i,j,m}\omega_{i,j,m}\dot{x}_{i,j,m}+\omega_{i,j,m}^2x_{i,j,m}=b_{i,j,m}u_m
\end{equation} 
with $i ={1,\ldots,n_{j,m}}$ denoting the considered natural frequency.
The state-space can be written as
\begin{subequations}
	\begin{align}
		\dot{\boldsymbol{x}}_{j,m}(t) = \boldsymbol{A}_{j,m} \, \boldsymbol{x}_{j,m}(t)+\boldsymbol{B}_{j,m} \, u_m(t)\\
		\boldsymbol{y}_{j,m}(t)=\boldsymbol{C}_{j,m} \, \boldsymbol{x}_{j,m}(t)
	\end{align}
\end{subequations}
with
\begin{subequations}
	\begin{align}
		\boldsymbol{x}_{j,m} &=
		\begin{pmatrix}
			x_{1,j,m} & x_{2,j,m} & \cdots & x_{n_{j,m},j,m}
		\end{pmatrix},\\
		\boldsymbol{A}_{i,j,m} &= \begin{pmatrix}
			0&1\\
			-\omega_{i,j,m}^2&-d_{i,j,m}\omega_{i,j,m}
		\end{pmatrix},\\
		\boldsymbol{A}_{j,m} &=
		\text{diag}\begin{pmatrix}
			\boldsymbol{A}_{1,j,m} & \cdots & \boldsymbol{A}_{n_{j,m},j,m}
		\end{pmatrix},\\
		\boldsymbol{B}_{j,m} &= 
		\begin{pmatrix}
			0 & b_{1,j,m} & \cdots & 0& b_{n_{j,m},j,m}
		\end{pmatrix}\tran\\
		\boldsymbol{C}_{j,m} &= 
		\begin{pmatrix}
			1 & 0 & \cdots & 1 & 0
		\end{pmatrix}.
	\end{align}
\end{subequations}
The input $u_m$ is found by the wind model in Section~\ref{ssec:wind_model}.
The parameters $b_{i,j,m}$, which essentially scale the input for each mode based on its dominance, cannot be directly identified without knowledge of the input.
However, since the input cannot be measured, it is assumed to be identical for all mirrors, their degrees of freedom, and the corresponding modes.
As a result, the parameters $b_{i,j,m}$ can only be identified relative to one another.
To obtain realistic simulation results, the overall input must be scaled appropriately.

\begin{table}
	\centering
	\vspace{1.5mm}
	\begin{tabular}{r c ccc c c c:c c}
		\toprule
		&& \multicolumn{3}{c}{M1} && M3 & \multicolumn{2}{c}{ } & M2 \\
		\cmidrule(lr){3-5} \cmidrule(lr){7-7} \cmidrule(lr){10-10}
		&& Piston & Tip & Tilt && Piston &&& Piston \\[0.4em]
		
		$\omega_1$ & & 22.9 & & & &   &&& 40.1\\	
		$d_1$ && 0.13 &  & & &  &&& 0.097\\	
		$b_1$ && 0.8 &  & & & &&& 12.2 \\[0.4em]
		
		$\omega_2$ && 24.3 &  & & &  24.2&&& 42.3 \\	
		$d_2$ && 0.13 &  &  & &  0.087&&&0.014 \\	
		$b_2$ && 0.7 &  & & & 1.8&&& 1.6  \\[0.4em]
		
		$\omega_3$ && 32.2 &  & 32.2 && 32.1 &&& 43.3 \\	
		$d_3$ && 0.13 &  & 0.12 &&  0.092 &&&0.015 \\	
		$b_3$ && 4.0 &  & 5.0 & & 8.63&&& 2.0 \\[0.4em]
		
		$\omega_4$ &&  &  & 33.3 & & 33.2 &&& 46.7\\
		$d_4 $&&  &  & 0.16 && 0.087 &&&  0.015\\
		$b_4$ &&  &  & 4.67 && 4.0 &&& 2.05 \\[0.4em]
		
		$\omega_5$ && 40.1 &  40.1 & 40.1&&  40.1 &&& 47.1\\	
		$d_5$ && 0.087 & 0.070 & 0.14 &&  0.12  &&& 0.016 \\	
		$b_5$ && 19 & 0.689 & 4.94 && 13.9&&&  3.1 \\[0.4em]
		
		$\omega_6$ && 48.0 & 48.1 & 51.3 && 51.2 &&& 50.5  \\	
		$d_6$ && 0.12 & 0.12 & 0.058 &&  0.058  &&&0.017 \\	
		$b_6$ && 12 & 2.5 & 1.5&& 2.75 &&&  22.5\\[0.4em]
		
		$\omega_7$ && 56.0 &  &52.6 && &&& 50.9 \\	
		$d_7$ && 0.49 &  & 0.098 && &&&  0.016 \\	
		$b_7$ && 90 &  & 4.1 & &&&&  6.3\\[0.4em]
		
		$\omega_8$ && 71.7 &  71.9 & 71.2 & & 71.2&&&  \\	
		$d_8$ && 0.35 & 0.29 & 0.55  && 0.46&&&  \\	
		$b_8$ && 59 & 50 & 195 &&  400&\multicolumn{2}{c}{ }&\\[0.4em]
		\bottomrule
	\end{tabular}
	\caption{Identified parameters for the dynamic mirror model.}
	\label{tab:ident}
\end{table}
\begin{figure*}
	\centering
	\vspace{0.5mm}
	\input{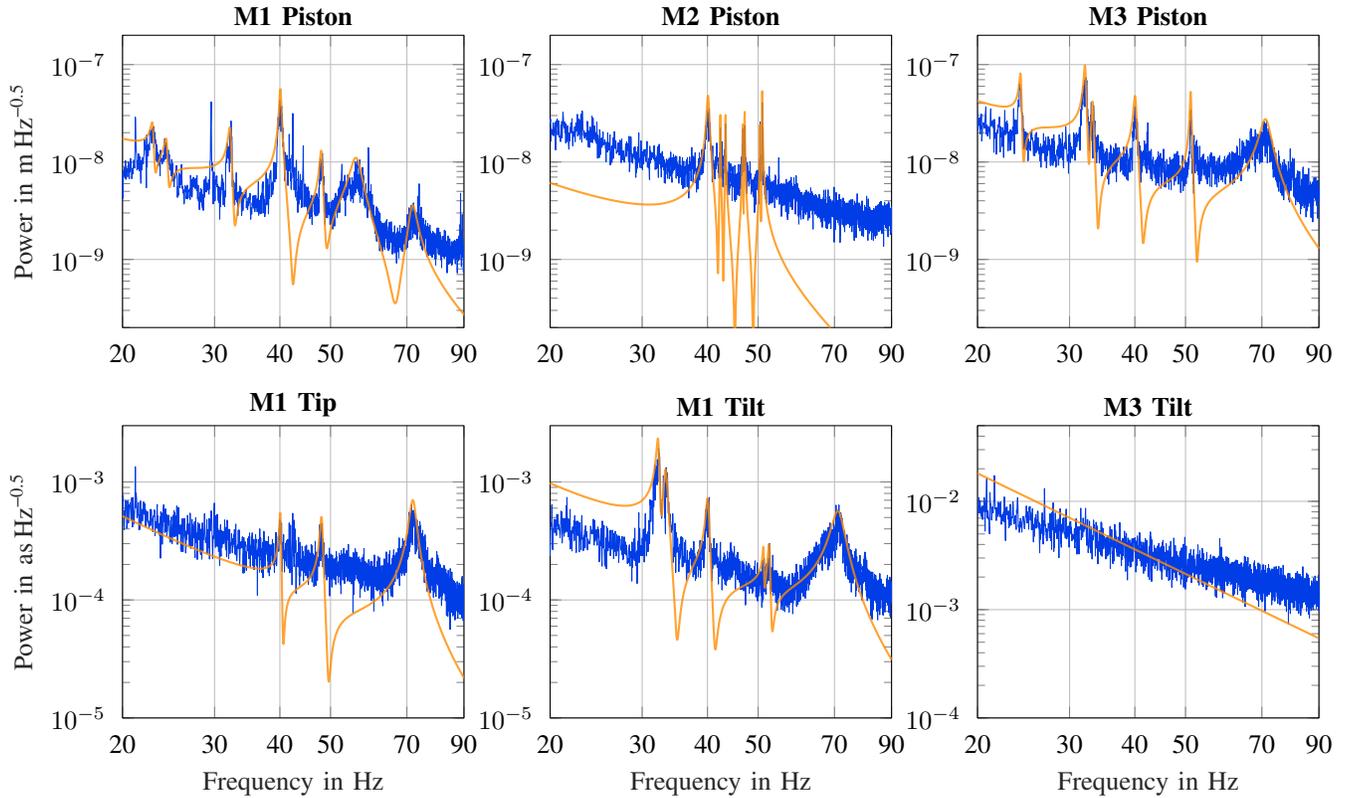}
	% Einfuegen nach xticks :
	% xticklabels={20, 30, 40, 50, 70, 90},
	% Einfuegen nach color=mycolor2, :
	% opacity=0.8,
	\definecolor{dark orange 70}{RGB}{255, 175, 77}
	\caption{Identification Results. \acs{PSD} of the mirror modes \leg{NGS}, identified wind and mirror model \leg{dark orange 70}.}%
	\label{fig:PSD_MMM1}%
\end{figure*}

%------------------------------------------------------------------------------------------------
\section{Identification results}\label{sec:results}
We provide the identification results using the aforementioned dynamical mirror model.
The identified parameters for the \ac{VLT} mirrors are visible in Table~\ref{tab:ident}.
For \ac{M3}, only eigenfrequencies of the piston mode are given, as the power spectrum of \ac{M3} tilt does not show any major peaks.
Certain frequencies, such as \qty{40.1}{\Hz}, appear in more than one mirror mode.
However, not all of the identified frequencies are observed in all measured modes.
This may be due to asymmetrical mechanical geometries at the mirror or uneven wind buffeting.
Specifically, the \qty{40.1}{\Hz} frequency is present in all modes suggesting a resonance of the entire telescope structure.
Likewise, all identified frequencies of the \ac{M3} piston (24, 32, 33, 51 and \qty{71}{\Hz}) are also found in the \ac{M1} modes.
This indicates a vibration of the lower telescope frame, where the \ac{M1} and \ac{M3} mirrors are mounted.

The \ac{PSD} of the identified mirror modes is detailed in Figure~\ref{fig:PSD_MMM1}.
The key findings of the identification can be summarized as follows.
The simulation successfully replicates the major eigenfrequencies of the mirrors.
The broad peaks in the spectrum, corresponding to structural vibrations induced by wind and other external forces, closely align with the measured accelerometer data.
This indicates that the model captures the dominant dynamic behavior of the system.
The narrow peaks arising from electrical machines or other causes are neglected by the model as it focuses only on the mechanical disturbances (c.f. Figure~\ref{fig:M1_piston_PSD_ex}).
The combination of the von Karman wind model and the identified mirror dynamics yield a reconstructed response closely matching the measured accelerometer signals.
While the model is accurate in the \qtyrange{20}{75}{\Hz} band, some discrepancies are observed beyond this range.
Specifically, the model exhibits a steeper slope for higher frequencies.
This is due to the PT2 characteristic of the identified mass-spring-damper systems, which induce an additional roll-off of \qty{-40}{\decibel} per decade.
However, this effect is not considered critical, as atmospheric effects dominate at lower frequencies, while measurement noise becomes significant at higher frequencies.
%
%Specifically, due to the inherent properties of double integration and the PT2 dynamics of the model, the slope of the simulated spectrum diverges from the experimental data outside of our considered range.
%
%It should be noted that the identification is only effective and feasible based on mass-spring-damper systems in a predefined frequency range that is not too wide.
%
%An analysis of the frequency domain of the model and the data shows that the slope of the the von Karman wind model and the accelerometer data approximately match for all frequencies.
%
%The model shows a steeper slope for higher frequencies though.
%
%This is caused by the PT2 characteristic of the identified model, which induces an additional roll-off of -40 dB per decade.

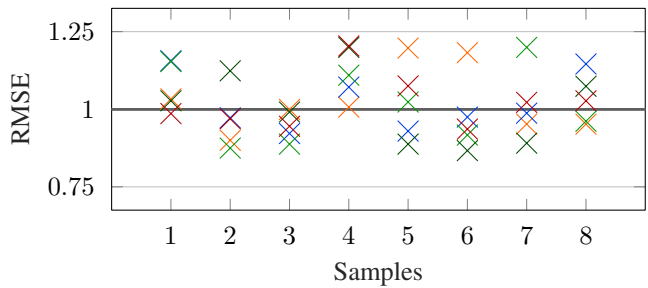
\begin{figure}
	\centering
	\vspace{2mm}
\definecolor{mycolor1}{rgb}{1.00000,0.40392,0.00000}%
\definecolor{mycolor2}{rgb}{0.37647,0.37647,0.37647}%
\definecolor{mycolor3}{rgb}{0.00000,0.23922,0.90196}%
\definecolor{mycolor4}{rgb}{0.03922,0.58824,0.03922}%
\definecolor{mycolor5}{rgb}{0.00000,0.29412,0.00000}%
\definecolor{mycolor6}{rgb}{0.70588,0.00000,0.00000}%
\begin{tikzpicture}

\begin{axis}[%
width=2.793in,
height=1.057in,
at={(0.525in,0.418in)},
scale only axis,
xmin=0,
xmax=9,
xtick={1, 2, 3, 4, 5, 6, 7, 8},
xlabel style={font=\color{white!15!black}},
xlabel={Samples},
ymin=0.675,
ymax=1.325,
ytick={0.75,    1, 1.25},
ylabel style={font=\color{white!15!black}},
ylabel={RMSE},
axis background/.style={fill=white},
ymajorgrids,
yticklabel style={xshift = -1mm,/pgf/number format/fixed},
xticklabel style={yshift = -1mm,/pgf/number format/fixed},
/tikz/line join=round,
title style={yshift=-6pt}
]
\addplot[only marks, mark=x, mark options={}, mark size=5.5902pt, draw=mycolor1, forget plot] table[row sep=crcr]{%
x	y\\
1	1.0334904865576\\
2	0.900057043937614\\
3	0.999330125123649\\
4	1.00686802452496\\
5	1.19707947604871\\
6	1.18230998042962\\
7	0.952351360944333\\
8	0.951951644008983\\
};
\addplot [color=mycolor2, line width=0.8pt, forget plot]
  table[row sep=crcr]{%
0	1\\
9	1\\
};
\addplot[only marks, mark=x, mark options={}, mark size=5.5902pt, draw=mycolor3, forget plot] table[row sep=crcr]{%
x	y\\
1	1.15621427691951\\
2	0.972219422552088\\
3	0.92191814541142\\
4	1.07139869809189\\
5	0.92967409681087\\
6	0.974822167235741\\
7	0.987382954824732\\
8	1.1458358805311\\
};
\addplot [color=mycolor2, line width=0.8pt, forget plot]
  table[row sep=crcr]{%
0	1\\
9	1\\
};
\addplot[only marks, mark=x, mark options={}, mark size=5.5902pt, draw=mycolor4, forget plot] table[row sep=crcr]{%
x	y\\
1	1.15355526949921\\
2	0.874861301341588\\
3	0.887762567211355\\
4	1.10965651799552\\
5	1.02329149708874\\
6	0.916927041288126\\
7	1.19921015749569\\
8	0.961464446732347\\
};
\addplot [color=mycolor2, line width=0.8pt, forget plot]
  table[row sep=crcr]{%
0	1\\
9	1\\
};
\addplot[only marks, mark=x, mark options={}, mark size=5.5902pt, draw=mycolor5, forget plot] table[row sep=crcr]{%
x	y\\
1	1.02721168227044\\
2	1.12373602695059\\
3	0.991012043533496\\
4	1.19923535928781\\
5	0.887727280017248\\
6	0.866777217858937\\
7	0.890771397115566\\
8	1.07406217995407\\
};
\addplot [color=mycolor2, line width=0.8pt, forget plot]
  table[row sep=crcr]{%
0	1\\
9	1\\
};
\addplot[only marks, mark=x, mark options={}, mark size=5.5902pt, draw=mycolor6, forget plot] table[row sep=crcr]{%
x	y\\
1	0.985997753510171\\
2	0.969364055731024\\
3	0.945238332608373\\
4	1.20384902068798\\
5	1.07514769516745\\
6	0.936130498198036\\
7	1.02225211699875\\
8	1.0270696273237\\
};
\addplot [color=mycolor2, line width=0.8pt, forget plot]
  table[row sep=crcr]{%
0	1\\
9	1\\
};
\end{axis}
\end{tikzpicture}%
	\caption{Temporal stability of identified model.
		\acs{RMSE} of M1 piston (\textcolor{LGS}{$\times$}),
		M1 tip (\textcolor{NGS}{$\times$}),
		M1 tilt (\textcolor{GPRoff}{$\times$}),
		M2 piston (\textcolor{GPRon}{$\times$}),
		M3 piston (\textcolor{Butter}{$\times$}).}%
	\label{fig:RMS_comp}%
\end{figure}

The temporal stability of the identified simulation model is investigated in Figure~\ref{fig:RMS_comp}.
By dividing the nighttime window of \qty{8}{\hour} into 8 segments of equal length, we can analyze the validity of the model for different conditions.
The quality of the identified model over a sample is found by calculating the \ac{RMSE} between the respective \ac{VLT} accelerometer data and the model from the \ac{PSD}.
The mirror model parameters are kept the same for all samples.
An average or overall \ac{RMSE} is also computed for the full night signals.
To visualize all mirror modes in a single chart, the computed \ac{RMSE} of each section and mode is divided by the overall \ac{RMSE}.
Thus, a value of 1 in Figure~\ref{fig:RMS_comp} indicates that the \ac{RMSE} of a sample is equivalent to the overall \ac{RMSE}.
Values greater than 1 imply that the identification error of a sample is larger, while values less than 1 suggest that the error is smaller than the \ac{RMSE} for the entire night.
As evident in Figure~\ref{fig:RMS_comp}, the error of all samples and mirror modes is within the range of \qty{\pm25}{\percent} compared to the overall \ac{RMSE}.
The identification is therefore not only valid for the average of a night but also for smaller subwindows, demonstrating  the robustness of the model.

% ------------------------------------------------------------------------------------------------
\section{Conclusion}\label{sec:conclusion}
% ---------- Summary/conclusion ----------
The results show that the identification of the individual mirror modes piston, tip and tilt based on accelerometer measurements is feasible and valid within a predefined frequency range.
The broad peaks in the spectrum, corresponding to wind and mechanical excitation, are accurately represented by the model, while the narrow peaks, mainly caused by electrical machines, are intentionally disregarded.
This is due to the focus of the model on structural dynamics rather than electronic interference sources.
The proposed mass-spring-damper system provides a robust approximation of the mirror dynamics within the considered frequency range of \qtyrange{20}{90}{Hz}.
The incorporation of a von Karman wind spectrum proved essential for identifying the model parameters and subsequently fitting the model to the measurement data.
%
%By dividing the accelerometer data by the wind spectrum, the dynamic mirror response was successfully isolated.
%
Beyond the range considered and for the wind noise floor outside the peaks, the accuracy of the model decreases.
It is expected, that improvements to the structure of the model, e.g. an modification of the spring-mass-damper differential equation, will further refine the identification for these regions.
%However, this limitation is not overly restrictive, as the relevant frequency range can be predicted in advance for different telescopes.
%
In general, the capacity to accurately identify the telescope dynamics was significantly impeded by the sparse distribution of accelerometers across the \ac{VLT} mirrors.
%
%Particularly \ac{M2}, and to a lesser extent, \ac{M3}.
%
To improve the mirror mode reconstruction from accelerometer data, at least four sensors should be placed on each mirror in the optical path.
It is recommended that the dominating mirror vibration modes should be considered when designing new acceleration measurement setups.
In summary, the findings of this study provide a feasible approach to modeling mirror vibrations in future large telescopes.
This especially enables model-based controllers such as \acs{LQG} which has been successfully implemented in simulation and at the \acl{GTC} \cite{Glueck2017, Marquis2024}.
%The findings of this study provide a simple and feasible approach to modeling mirror vibrations in future telescopes, such as the \ac{ESO} \ac{ELT}.

% ---------- Outlook ----------
The developed methods will contribute to the mitigation of vibration-induced disturbances in \ac{AO} systems, improving the accuracy of observations.
Future research should focus on validating the model through targeted excitation experiments at the telescope.
%
% Further research should investigate extending the model to account for coupling effects between the mirrors and the telescope structure as a whole.
%
Moreover, the integration of \ac{WFS} data with accelerometer measurements has the potential to further improve the reconstruction of mirror oscillations.
By addressing these aspects, the proposed technique can provide a more accurate model of the vibrations, alleviating the need for costly and complex black box models.

% ------------------------------------------------------------------------------------------------
\section*{Acknowledgment}
The authors would like to thank the German Federal Ministry of Education and Research (BMBF) for supporting this work under grant 05A23VS1 as well as the colleagues at Institute for System Dynamics (University of Stuttgart) and Max Planck Institute for Astronomy for their helpful suggestions and comments.

% ------------------------------------------------------------------------------------------------
% References
\bibliographystyle{IEEEtranS} % makes bibtex use spiebib.bst
%\bibliography{bibliography_PJ_OneAuthor} % bibliography data in bibliography_PJ.bib
\bibliography{bibliography_PJ} % bibliography data in bibliography_PJ.bib

% Generated by IEEEtranS.bst, version: 1.14 (2015/08/26)
\begin{thebibliography}{10}
\providecommand{\url}[1]{#1}
\csname url@samestyle\endcsname
\providecommand{\newblock}{\relax}
\providecommand{\bibinfo}[2]{#2}
\providecommand{\BIBentrySTDinterwordspacing}{\spaceskip=0pt\relax}
\providecommand{\BIBentryALTinterwordstretchfactor}{4}
\providecommand{\BIBentryALTinterwordspacing}{\spaceskip=\fontdimen2\font plus
\BIBentryALTinterwordstretchfactor\fontdimen3\font minus
  \fontdimen4\font\relax}
\providecommand{\BIBforeignlanguage}[2]{{%
\expandafter\ifx\csname l@#1\endcsname\relax
\typeout{** WARNING: IEEEtranS.bst: No hyphenation pattern has been}%
\typeout{** loaded for the language `#1'. Using the pattern for}%
\typeout{** the default language instead.}%
\else
\language=\csname l@#1\endcsname
\fi
#2}}
\providecommand{\BIBdecl}{\relax}
\BIBdecl
\renewcommand{\BIBentryALTinterwordstretchfactor}{4}

\bibitem{Bely2003}
\BIBentryALTinterwordspacing
P.~Y. Bely, \emph{{The Design and Construction of Large Optical Telescopes}},
  1st~ed., ser. Astronomy and Astrophysics Library.\hskip 1em plus 0.5em minus
  0.4em\relax New York, NY: Springer, Jan. 2003. \relax
  \url{https://doi.org/10.1007/b97612}
\BIBentrySTDinterwordspacing

\bibitem{Boehm2013}
\BIBentryALTinterwordspacing
M.~B{\"o}hm, J.-U. Pott, M.~Kürster, and O.~Sawodny, ``{Modeling and
  Identification of the Optical Path at ELTs - a Case Study at the LBT},''
  \emph{IFAC Proceedings Volumes}, vol.~46, no.~5, pp. 249--255, 2013, 6th IFAC
  Symposium on Mechatronic Systems. \relax
  \url{https://www.sciencedirect.com/science/article/pii/S1474667015362236}
\BIBentrySTDinterwordspacing

\bibitem{Boehm2012}
\BIBentryALTinterwordspacing
M.~B{\"o}hm, T.~Ruppel, J.-U. Pott, O.~Sawodny, T.~Herbst, and M.~K{\"u}rster,
  ``{Modelling the optical pathway of the Large Binocular Telescope},'' in
  \emph{Modeling, Systems Engineering, and Project Management for Astronomy V},
  G.~Z. Angeli and P.~Dierickx, Eds., vol. 8449, International Society for
  Optics and Photonics.\hskip 1em plus 0.5em minus 0.4em\relax SPIE, 2012, p.
  844915. \relax \url{https://doi.org/10.1117/12.926293}
\BIBentrySTDinterwordspacing

\bibitem{Enard1989}
\BIBentryALTinterwordspacing
D.~Enard and F.~Merkel, ``{The Optics Of The Very Large Telescope},'' in
  \emph{Optical Design Methods, Applications and Large Optics}, A.~Masson,
  J.~J.~S. in-den Baeumen, and H.~Zuegge, Eds., vol. 1013, International
  Society for Optics and Photonics.\hskip 1em plus 0.5em minus 0.4em\relax
  SPIE, 1989, pp. 192 -- 197. \relax \url{https://doi.org/10.1117/12.949378}
\BIBentrySTDinterwordspacing

\bibitem{Gilbert2010}
H.~B. Gilbert, O.~Celik, and M.~K. O'Malley, ``Long-term double integration of
  acceleration for position sensing and frequency domain system
  identification,'' in \emph{2010 IEEE/ASME International Conference on
  Advanced Intelligent Mechatronics}.\hskip 1em plus 0.5em minus 0.4em\relax
  IEEE, 2010, pp. 453--458.

\bibitem{Glueck2017}
M.~Gl{\"{u}}ck, J.-U. Pott, and O.~Sawodny, ``{I}nvestigations of an
  {A}ccelerometer-based {D}isturbance {F}eedforward {C}ontrol for {V}ibration
  {S}uppression in {A}daptive {O}ptics of {L}arge {T}elescopes,''
  \emph{Publications of the Astronomical Society of the Pacific}, vol. 129,
  Apr. 2017.

\bibitem{Hovsepian1998}
\BIBentryALTinterwordspacing
T.~Hovsepian, J.~L. Michelin, and S.~Stanghellini, ``{Design and tests of the
  VLT M1 mirror passive and active supporting system},'' in \emph{Advanced
  Technology Optical/IR Telescopes VI}, L.~M. Stepp, Ed., vol. 3352,
  International Society for Optics and Photonics.\hskip 1em plus 0.5em minus
  0.4em\relax SPIE, 1998, pp. 424 -- 435. \relax
  \url{https://doi.org/10.1117/12.319264}
\BIBentrySTDinterwordspacing

\bibitem{Jaufmann2024}
\BIBentryALTinterwordspacing
P.~Jaufmann, A.~Buck, M.~Zaiser, J.-U. Pott, and O.~Sawodny, ``{Simulation of
  wavefront-based disturbance observers for large telescopes},'' in
  \emph{Adaptive Optics Systems IX}, K.~J. Jackson, Ed., vol. 13097,
  International Society for Optics and Photonics.\hskip 1em plus 0.5em minus
  0.4em\relax SPIE, 2024, p. 130977Q. \relax
  \url{https://doi.org/10.1117/12.3018415}
\BIBentrySTDinterwordspacing

\bibitem{Kulcsar2012a}
C.~Kulcs{\'a}r, P.~Massioni, G.~Sivo, and H.-F.~G. Raynaud, ``Vibration
  mitigation in adaptive optics control,'' in \emph{Adaptive Optics Systems
  {III}}, B.~L. Ellerbroek, Ed., vol. 8447.\hskip 1em plus 0.5em minus
  0.4em\relax {SPIE}, Sep. 2012, pp. 381--396.

\bibitem{Kulcsar2012b}
\BIBentryALTinterwordspacing
C.~Kulcs{\'a}r, G.~Sivo, H.-F. Raynaud \emph{et~al.}, ``{Vibrations in AO
  control: a short analysis of on-sky data around the world},'' in
  \emph{Adaptive Optics Systems III}, B.~L. Ellerbroek, Ed., vol. 8447,
  International Society for Optics and Photonics.\hskip 1em plus 0.5em minus
  0.4em\relax SPIE, 2012, p. 84471C. \relax
  \url{https://doi.org/10.1117/12.925984}
\BIBentrySTDinterwordspacing

\bibitem{MacMynowski2010}
D.~G. MacMynowski and T.~Andersen, ``Wind buffeting of large telescopes,''
  \emph{Applied optics}, vol.~49, no.~4, pp. 625--636, 2010.

\bibitem{Marquis2024}
L.~Marquis, H.-F. Raynaud, N.~Galland \emph{et~al.}, ``{First on-sky tests of
  LQG control for a 10m-class telescope: prelude on the Gran Telescopio
  Canarias adaptive optics system},'' in \emph{Adaptive Optics Systems IX},
  vol. 13097.\hskip 1em plus 0.5em minus 0.4em\relax SPIE, 2024, pp.
  1869--1875.

\bibitem{Stanghellini1996}
S.~Stanghellini, ``The secondary mirror units of the vlt: design overview and
  manufacturing status.'' \emph{The Messenger, vol. 86, p. 5-8}, vol.~86, pp.
  5--8, 1996.

\end{thebibliography}
\end{document}